\documentclass[conference]{IEEEtran}
\IEEEoverridecommandlockouts
\usepackage[spanish]{babel}
\usepackage{cite}
\usepackage{amssymb}
\usepackage{amsmath,amsfonts}
\usepackage{algorithm}
\usepackage{algpseudocode}
\usepackage{graphicx}
\usepackage{textcomp}
\usepackage{xcolor}
\def\BibTeX{{\rm B\kern-.05em{\sc i\kern-.025em b}\kern-.08em
    T\kern-.1667em\lower.7ex\hbox{E}\kern-.125emX}}
\begin{document}

\title{Accelerating the Convex Hull Computation with a Parallel GPU Algorithm 
}

\newcommand{\cristobal}[1]{\textbf{\textcolor{magenta}{\{Cristobal: #1\}}}}

\author{\IEEEauthorblockN{1\textsuperscript{st} Alan Keith}
\IEEEauthorblockA{Instituto de Informática\\
Universidad Austral de Chile\\
Valdivia, Chile \\
alan.keithpaz@gmail.com}
\and
\IEEEauthorblockN{2\textsuperscript{nd} Héctor Ferrada}
\IEEEauthorblockA{Instituto de Informática\\
Universidad Austral de Chile\\
Valdivia, Chile \\
hferrada@inf.uach.cl}
\and
\IEEEauthorblockN{3\textsuperscript{rd} Cristóbal A. Navarro}
\IEEEauthorblockA{Instituto de Informática\\
Universidad Austral de Chile\\
Valdivia, Chile \\
cnavarro@inf.uach.cl}
}

\maketitle

\begin{abstract}

The convex hull is a fundamental geometrical structure for many applications where groups of points must be enclosed or represented by a convex polygon. Although efficient sequential convex hull algorithms exist, and are constantly being used in applications, their computation time is often considered an issue for time-sensitive tasks such as real-time collision detection, clustering or image processing for virtual reality, among others, where fast response times are required. 
In this work we propose a parallel GPU-based adaptation of heaphull, which is a state of the art CPU algorithm that computes the convex hull by first doing a efficient filtering stage followed by the actual convex hull computation. More specifically, this work parallelizes the filtering stage, adapting it to the GPU programming model as a series of parallel reductions. Experimental evaluation shows that the proposed implementation significantly improves the performance of the convex hull computation, reaching up to $4\times$ of speedup over the sequential CPU-based heaphull and between $3\times \sim 4\times$ over existing GPU based approaches.

\end{abstract}

\begin{IEEEkeywords}
Convex hull, GPU Computing, Filtering, Parallel Reduction, Warp Shuffle
\end{IEEEkeywords}

\section{Introducción}


El cómputo de la cerradura convexa, o \textit{convex hull} en ingles, es una tarea fundamental y de suma importancia para diversas aplicaciones de la ciencia y tecnología, como por ejemplo en geometría computacional, gráfica computacional y modelamiento de formas, entre otras. El convex hull es requerido frecuentemente para detección de colisiones, cálculo de interferencias, análisis de formas, reconocimiento de patrones, generación de estadísticas, sistemas de información geográfica, entre otras muchas disciplinas~\cite{barber_1996, Rourke_1985, Lin_1996}.

\begin{figure}[ht!]
\centerline{\includegraphics[scale=0.6]{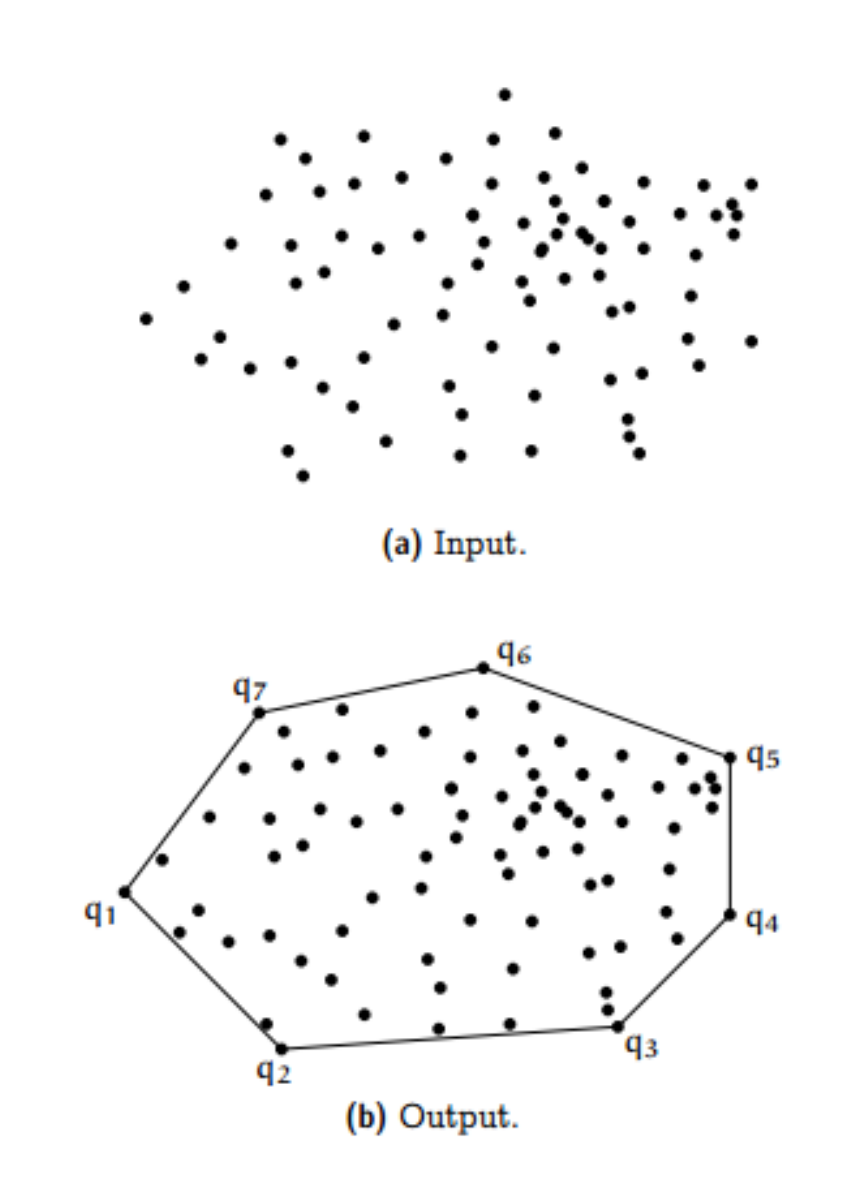}}
\caption{Cálculo del Convex Hull} \label{fig1}
\end{figure}

La convexidad es una propiedad relevante para variadas ramas de la ciencia que requieren de análisis geométrico. Una cerradura, que contiene un conjunto de puntos, es convexa si para todo par de puntos pertenecientes a esta, el segmento de línea recta que los une está completamente dentro de ella~\cite{cormen_2001}. Uno de los primeros problemas identificados en el campo de la geometría computacional es el de calcular la forma convexa más pequeña, llamada convex hull, que encierra un conjunto de puntos~\cite{preparata_1985}. Así, dado un conjunto de puntos en un espacio en $d$ dimensiones, el convex hull es su mínimo subconjunto convexo que contiene todos los puntos, tal como se muestra en la Figura \ref{fig1}.  
En otras palabras, dado un conjunto de puntos $S$, definimos que $CH(S)$ es la envolvente o cierre convexo de $S$ si i) $CH(S)$ es convexo, ii) $S$ está contenido en $CH(S)$ y iii) $CH(S)$ es el menor conjunto convexo que contiene a $S$. Para el caso de 2 dimensiones ($2D$), el convex hull corresponde al polígono convexo más pequeño que contiene a todos los puntos del conjunto. Intuitivamente podemos definir al convex hull como ``si $S$ es un conjunto finito de puntos en el plano e imaginamos rodear el conjunto con una banda elástica; cuando se suelte el elástico este asumirá la forma del convex hull''~\cite{preparata_1985}.

El cómputo del convex hull suele realizarse directamente sobre los puntos de entrada. Sin embargo, en muchos casos --- dependiendo de la instancia de entrada --- es conveniente primero realizar un filtrado de puntos, descartando elementos que son evidentemente innecesarios para el proceso final. Lo anterior da lugar a las siguientes dos etapas: 
\begin{enumerate}
    \item \textbf{Filtrado}. Este paso consiste en descartar el máximo número de puntos que evidentemente no pertenecen al convex hull, por medio de una validación matemática de bajo costo computacional; esto incluye a puntos que están sobre alguno de los lados de este polígono y no son vértices del segmento de recta.
    \item \textbf{Convex hull}. Recibe como entrada los puntos que han pasado el filtro de la etapa anterior y calcula el convex hull con algún algoritmo tradicional. De esta forma, el posterior cálculo del convex hull es realizado con un número reducido de puntos, siempre y cuando el paso anterior haya logrado descartar algunos puntos, acelerando y simplificando la selección de puntos del convex hull. El cálculo en sí es realizado por algoritmos conocidos como Graham Scan~\cite{graham_1972}, Jarvis March~\cite{jarvis_1973} u otro.
\end{enumerate}
\subsection{Algoritmos para el cómputo del convex hull}
El problema de calcular el convex hull de un conjunto de puntos ha sido ampliamente estudiado en la geometría computacional. Se han propuesto muchos algoritmos teóricamente óptimos para conjuntos de puntos de baja y alta dimensión. Este fue uno de los primeros problemas en el campo a ser estudiado desde el punto de vista de la complejidad computacional, determinando que el tiempo asintótico óptimo para todo algoritmo basado en comparaciones dentro del modelo RAM\footnote{Bajo el modelo de costos RAM, es posible realizar cualquier cálculo matemático simple entre dos palabras de memoria, de tamaño $O(\log n)$ bits, en tiempo constante.}, es de es $O(n \log n)$, dónde $n$ es la cantidad de puntos de entrada~\cite{cormen_2001,akl_1984}. Además, mediante un proceso lineal, es posible reducir el problema de ordenar $n$ puntos a encontrar el convex hull en este mismo conjunto; confirmando que la envolvente convexa se puede calcular en un tiempo óptimo de $O(n \log n)$, ya que esta es la cota inferior al problema de ordenamiento.
La Tabla \ref{tabla:tiempos_algoritmos} muestra los tiempos asintóticos de los algoritmos de convex hull históricamente más conocidos para dos dimensiones. Consideremos $n$ el número de puntos en el conjunto de entrada, $h$ el número de vértices del polígono de salida. Nótese que $h \le n$, por lo que $nh \le n^2$.

\begin{table}[htbp]
\caption{Tiempos de algoritmos para el cálculo del convex hull en $2D$.}
\begin{center}
\begin{tabular}{|l|l|l|}
\hline
\textbf{Algoritmo} & \textbf{Tiempos} & \textbf{Autor} \\
\hline
Fuerza Bruta & $O( n^4 )$ & Anónimo \\ \hline
Gift Wrapping & $O ( nh ) $ & Chand \& Kapur, 1970 \\ \hline
Graham Scan & $O ( n \log n ) $ & Graham, 1972 \\ \hline
Jarvis March & $O ( nh ) $ & Jarvis, 1973 \\ \hline
QuickHull & $O ( nh ) $ & Eddy, 1977, Bykat, 1978 \\ \hline
Divide-and-Conquer & $O ( n \log n ) $ & Preparata \& Hong, 1977 \\ \hline
Monotone Chain & $O ( n \log n ) $ & Andrew, 1979 \\ \hline
Incremental & $O ( n \log n ) $ & Kallay, 1984 \\ \hline
Marriage-before-Conquest & $O ( n \log n ) $ & Kirkpatrick \& Seidel, 1986 \\ \hline
\end{tabular}
\label{tabla:tiempos_algoritmos}
\end{center}
\end{table}

\subsection{Algoritmo Heaphull}
El trabajo de Ferrada et al.~\cite{ferrada_2019} está basado en el desarrollo de una técnica de optimización que reduce el costo computacional para construir el convex hull. Este método preprocesa el conjunto de puntos de entrada, filtrando todos los puntos dentro de un polígono de ocho vértices en tiempo $O(n)$ y devuelve un conjunto reducido de puntos candidatos para encontrar el convex hull, los cuales se encuentran semi ordenados y distribuidos en cuatro colas de prioridad (una para cada cuadrante).

Los resultados experimentales muestran que en el peor de los casos (cuando todos los puntos se encuentran sobre una circunferencia) un pequeño desplazamiento radial aleatorio de los puntos hace de este método el más rápido. Además, al aumentar la magnitud de este desplazamiento, el rendimiento del método propuesto escala a un ritmo más rápido que las otras implementaciones testeadas. En términos de eficiencia de memoria, esta implementación logra usar de $3\times$ a $6\times$ menos memoria que otros métodos.

Este algoritmo es esencialmente una técnica de filtrado que se conforma de cuatro etapas relevantes, (1) construcción de un octágono con los puntos extremos, (2) filtrado de puntos y agrupación en colas de prioridad, (3) ordenamiento parcial y (4) cálculo del convex hull. Las primeras tres etapas toman tiempo $O(n)$, mientras que la etapa de cálculo del convex hull toma $O (n' \log n' )$, donde $n'$ es el tamaño del conjunto filtrado $P' \subseteq P$. El Algoritmo~\ref{alg_heaphull} muestra la idea de heaphull.

\begin{center}
\begin{algorithm}[H]
\small
\caption{\textbf{heaphull}, Ferrada et al.~\cite{ferrada_2019}. To compute the convex hull $CH[1..h]$ in $2D$}\label{euclid}
\begin{algorithmic}[1]
\Require $n$ points in $2D$, stored in an array $P[1..n]$.
\Ensure an array $CH[1..h]$ with the $h$ points of the hull in counter-clockwise order.
\Procedure{heaphull}{$P,n$}
\State $E\gets\textsc{findExtremes}($P$, $n$)${\scriptsize\Comment{{\color{blue}find extreme points ($O(n)$ time)}}}
\State $n_1=n_2=n_3=n_4=0${\scriptsize\Comment{{\color{blue}counter for the 4 priority queues $Q_i$}}}
\For{\texttt{$j=1$ to $n$}}
	\If{$P[j]$ is outside the convex octagon $CP(E)$}
		\State $i\gets\textsc{findQueue}(P,n,j$) {\scriptsize\Comment{{\color{blue}find $Q_i$ for $P[j]$}}}
		\State store $j$ in the priority queue $Q_i$ 
		\State $n_i\gets n_i+1$
	\EndIf
\EndFor
\State $CH\gets \emptyset$
\For{\texttt{$i=1$ to $4$}}
    \scriptsize\Comment{{\color{blue}add the CH for points of $Q_i$ ($O(n_i\log n_i)$ time)}}
	\State $CH\gets CH\cup\textsc{hull}(Q_i, n_i)$
\EndFor
\State \textbf{return} $CH$
\EndProcedure
\end{algorithmic}
\label{alg_heaphull}
\end{algorithm}
\end{center}

\subsection{Paralelismo GPU en el cómputo del Convex Hull}

Las GPU (Graphic Process Unit) se han utilizado para acelerar el rendimiento de diversas aplicaciones, como la simulación de partículas, el modelado molecular y el procesamiento de imágenes. La arquitectura GPU esta basada en el paralelismo masivo, proporcionando cientos de cores (elementos de procesamiento), donde cada core es un procesador pipeline multi-etapa. Los cores se agrupan para generar un multiprocesador (SM) de caracter SIMD al nivel de grupos de 32 threads. Cada SM tiene su propio conjunto de registros y memoria local, y los cores dentro del mismo SM tienen memoria compartida limitada. 


Este trabajo propone una adaptacion del algoritmo heaphull al modelo de programacion GPU, para acelerar su rendimiento y aprovechar el paralelismo de datos disponible en la etapa de filtrado. En particular se busca responder las siguientes preguntas:
\begin{itemize}
    \item ¿Cuánto mejora el desempeño de la etapa de filtrado al paralelizar en GPU el algoritmo heaphull?
    \item Al paralelizar heaphull ¿Se logra que el algoritmo sea más eficiente que otros que ya usan paralelismo?
\end{itemize}
El resto de este artículo se divide en 5 secciones. En la primera sección se describe el trabajo relacionado, considerando especialmente los algoritmos que realizan un esfuerzo por paralelizar la etapa de filtrado. Para esto se clasificaron en algoritmos secuenciales tradicionales, algoritmos secuenciales que explotan el filtrado previo, algoritmos paralelos de filtrado en CPU y algoritmos paralelos de filtrado en GPU. En la segunda sección se define en detalle la metodología de investigación implementada. En la sección 3 se presenta el diseño del algoritmo paralelo en CPU y en GPU. En la sección 4 se presentan los resultados de rendimiento (speedup, eficiencia y comparación entre CPU y GPU). Finalmente en la sección 5 se desarrollan las conclusiones.

\section{Trabajo Relacionado}
Existen varios métodos prácticos conocidos e implementaciones de software robustas que abordan el cómputo del convex hull de manera secuencial. Desde 1970 se han desarrollado varios algoritmos clásicos~\cite{jiayin_2015}, como Graham scan~\cite{graham_1972}, Gift wrapping~\cite{jarvis_1973}, Incremental method~\cite{kallay_1984}, Divide-and-Conquer~\cite{preparata_1977}, Monotone chain~\cite{andrew_1979}, and Quick-Hull~\cite{barber_1996}. 

\subsection{Convex hull con etapa de filtrado}
Realizar un filtrado eficiente del conjunto de puntos puede tener un gran impacto en el tiempo final de cálculo. Vyšniauskaitė~\cite{laura_2006} en 2006 propuso una idea de la filtración a priori de puntos, presentado dos nuevos algoritmos basados en la búsqueda de puntos extremos y en la subdivisión del conjunto de puntos en arreglos más pequeños.

También Sharif~\cite{sharif_2011} el 2011 propone un método híbrido para calcular el convex hull, basado en dos algoritmos ya existentes, es decir, Quickhull y GrahamScan, intentando eliminar las deficiencias en las dos técnicas mencionadas anteriormente usando el primero para realizar un filtrado y realizando el cálculo final con el segundo de estos.

\subsection{Convex hull paralelo}
Para mejorar la eficiencia computacional del convex hull se han hecho algunas contribuciones valiosas rediseñando e implementando algoritmos de convex hull secuenciales en paralelo, explotando el potencial de cálculo masivo de las GPU. La mayoría de estas implementaciones se han desarrollado teniendo como base el algoritmo QuickHull~\cite{jiayin_2015}. 
\subsubsection{Paralelismo en CPU}
Como se muestra la publicación de Chen y otros~\cite{chen_1992} el convex hull puede ser calculado de forma paralela en el modelo PRAM (Máquina de acceso aleatorio paralelo), siendo el algoritmo directamente implementable en CPUs multicore.

En 1987 Goodrich~\cite{goodrich_1987} diseñó un algoritmo paralelo para encontrar el convex hull que se ejecuta en tiempo $O(log ~n)$ utilizando  $O (n / log ~n)$ procesadores en el modelo computacional CREW (concurrent read and exclusive write) PRAM, que es óptimo. Una de las técnicas que utilizan para alcanzar estos límites óptimos es el uso de una estructura de datos paralela que llaman ``hull tree''.

Miller~\cite{miller_1988} en 1988 presentó algoritmos paralelos para identificar los puntos extremos del convex hull, concentrándose en el desarrollo de estos en tiempo polilogarítmico para una variedad de máquinas paralelas y analizándolos usando notación O. Asimismo, Liu~\cite{liu_2015} en 2015 hace una revisión completa de los algoritmos paralelos existentes a la fecha, categorizándolos según su arquitectura y considerando sus tiempos de ejecución en notación O y el número de procesadores necesarios para esto.

Berkman~\cite{berkman_1996} en 1996 desarrolló un algoritmo paralelo para encontrar el convex hull en tiempo $O (log ~log ~n)$ usando $n/log ~log ~n$ procesadores en una arquitectura PRAM CRCW (Concurrent read concurrent write) común. Para romper la barrera de tiempo $\Omega (log ~n / log ~log ~n)$ requerida para generar el convex hull, introduce una estructura de datos, un árbol balanceado de altura doblemente logarítmica y usa este para representarlo. El algoritmo demuestra el poder del ``paradigma doblemente logarítmico de divide y vencerás''.

Nakagawa~\cite{nakagawa_2009} presentó una implementación de algoritmo paralelo simple para calcular el convex hull y evaluar el rendimiento en los procesadores de cuatro núcleos duales, logrando un factor de aceleración de aproximadamente 7 utilizando 8 procesadores. Como el factor de aceleración de más de 8 no es posible, su implementación paralela para calcular el convex hull es casi óptima.

Waghmare~\cite{waghmare_2010} presenta un algoritmo de convex hull usando clustering K-means, en el que los puntos en 2D se agrupan en clusters diferentes y luego se calculan los convex hull para cada uno de estos. El algoritmo se implementa en modo MPI, OpenMP e híbrido. Los resultados indican que el enfoque híbrido supera al enfoque MPI y OpenMP por separado.

\subsubsection{Paralelismo en GPU}
El trabajo relacionado en computación paralela para el cálculo del convex hull en GPU muestra que se puede alcanzar mejor velocidad y rendimiento, como se evidencia en el trabajo de Tang et al.~\cite{tang_2012}, y Jiayin et al~\cite{jiayin_2015}. Estos algoritmos basados en GPU trabajan sobre la optimización del filtrado de puntos, realizando la etapa posterior de cálculo en forma secuencial.

Srungarapu et al.~\cite{srung_2011} propuso en 2011 una implementación optimizada en GPU para encontrar el convex hull para conjuntos de puntos de dos dimensiones. Su implementación trata de minimizar el impacto de los patrones de acceso irregular a los datos, logrando un speedup de hasta 14 veces sobre la implementación estándar secuencial en CPU.

En 2011 Jurkiewicz y Danilewski~\cite{jurki_2011} presentaron una propuesta de implementación de algoritmo de cálculo del convex hull implementada en su totalidad en GPU, sin filtrado y basado en 3 algoritmos secuenciales distintos, con pruebas experimentales con conjuntos de entrada de $10^7$ puntos, obteniendo resultados en reducción de tiempo de hasta $100$ veces para el máximo de puntos al compararse con una librería que utiliza algoritmos similares, pero que su desempeño se encuentra muy por debajo de la media.

Mei G.~\cite{mei_2016} en 2016 propuso una versión paralela en CUDA (GPU) para el cálculo de convex hull para dos~\cite{mei_2017} y tres~\cite {mei_2015} dimensiones, alcanzando aceleraciones de aproximadamente 4x en promedio y $5x \sim 6x$ en los mejores casos, la cual llama CudaChain. Realizando solo una mejora de filtrado, afirma que el 95\% de puntos de entrada se pueden descartar en la mayoría de las pruebas experimentales. Anterior a esto, usó GPUs para mejorar la fase de filtrado en la búsqueda de puntos extremos sobre datos sintéticos y con el uso de la biblioteca qhull (implementación de Quickhull en C++), la idea básica era descartar los puntos que se ubican dentro de un polígono convexo formado por 16 extremos~\cite{mei_2014}.

Por otro lado, Stein~\cite{stein_2012} en 2012 hizo una versión paralela para el convex hull en 3 dimensiones que se basa en el enfoque QuickHull y comienza construyendo un tetraedro inicial utilizando cuatro puntos extremos, descarta los puntos internos y distribuye los puntos externos a las cuatro caras y continúa luego iterativamente. En su trabajo, Stein afirma que su implementación superó al Qhull basado en CPU en 30 veces para 10 millones de puntos y 40 veces para 20 millones de puntos.


En 2020 Masnadi~\cite{masnadi_2020} presenta ConcurrentHull, una nueva técnica basada en el filtrado de puntos para conjuntos de datos 2D y 3D. Su implementación, que es una combinación de filtrado, divide y vencerás, y computación paralela, permite ser empleada en un entorno de computación distribuida. Su algoritmo tiene una versión para CPU y otra para GPU (CUDA). Los resultados muestran que tiene una ganancia de rendimiento con grandes tamaños de datos de entrada y tiene la ventaja de poder manejar grandes conjuntos de datos.

Si bien las soluciones propuestas informan un speedup importante para las condiciones establecidas para las pruebas, nuestro trabajo busca no solo mejorar los tiempos de ejecución si no ponemos especial atención en el porcentaje de filtrado a obtener en la fase paralela, lo cual tiene el mayor impacto en el tiempo de cálculo posterior. Nuestro trabajo compara las dos soluciones más actuales (CudaChain y ConcurrentHull) con la nuestra (GPU Heaphull), comparando la solución en el peor caso y en el caso promedio.

\section{Paralelización del Algoritmo Heaphull}
El primer paso de la versión paralela consiste en encontrar los cuatro puntos extremos en las direcciones norte, sur, este y oeste del conjunto de puntos y formar el cuadrilátero que definen. Tras este paso, se tienen puntos distribuidos en cuatro regiones, formadas por los puntos fuera del cuadrilátero. En cada uno de ellas existe un punto que es el más cercano a la esquina correspondiente. Tras la obtención de estos cuatro puntos adicionales, se obtiene un octágono que divide los puntos restantes externos en ocho regiones que se pueden tratar individualmente siguiendo la misma regla. 

La paralelización en GPU se realiza acelerando la reducción de mínimos y máximos para formar el polígono de filtrado, usando el enfoque de Shuffle Warp Reduce, el cual permite que conjuntos de 32 threads, llamados \textit{warps}, puedan comunicar sus registros de forma eficiente. 

A diferencia de la versión tradicional de heaphull, que implementa la búsqueda de los $8$ puntos en cada paso del bucle, en la versión GPU se ejecuta un \textit{kernel} de reducción para encontrar los primeros $4$ puntos extremos, seguido de un segundo \textit{kernel} de reducción para encontrar los $4$ puntos de las esquinas. La razón de por qué se deben ejecutar dos \textit{kernels} en sucesión es debido a que la búsqueda de los puntos secundarios requiere de los primeros $4$ puntos. 

En el segundo paso de la búsqueda de los puntos de esquina, se utilizan distancias de Manhattan, es decir, para cada esquina $c_i, 1 \leq i \leq 4$, se encuentra el punto en $P$ que minimiza la suma de las distancias vertical y horizontal a la esquina correspondiente. Si bien la distancia de Manhattan es una métrica que puede entregar puntos mas distantes a la esquina que la distancia Euclidiana, esto tiende a ocurrir con baja frecuencia, y cuando ocurre, se puede ver que el punto elegido todavía se encuentra muy cerca del más cercano por distancia Euclidiana~\cite{ferrada_2019}. La ventaja de usar la distancia Manhattan es que no requiere del cómputo de raíces cuadradas, lo cual lo convierte en una métrica mas rápida de calcular que la distancia euclidiana.

En el paso siguiente se realiza el descarte de los puntos que deben ser filtrados y se crean las colas. Para la creación de las 4 colas se usa un arreglo donde se asigna el número de cola ($1$, $2$, $3$ o $4$) a la posición donde se encuentra un punto que si pertenece a la cola respectiva y un $0$ cuando ha sido descartado. Luego, en la CPU se crean las 4 colas con los respectivos índices guardados para ser entregadas a la sección correspondiente del algoritmo secuencial heaphull.

El Algoritmo~\ref{alg_filterGPU} muestra los pasos computacionales del proceso.
\begin{center}
\begin{algorithm}
\small
\caption{\textbf{GPUfilter}. Para calcular el octágono de filtrado de $P[1..n]$ en $2D$}
\begin{algorithmic}[1]
\Require $n$ valores de punto flotante, en $2D$, almacenados en un arreglo $P[1..n]$.
\Ensure un arreglo $Q[1..n']$ con los índices de la cola a la que pertenece cada uno de los puntos.
\Procedure{GPUfilter}{$P,n$}
\State$E\gets\textsc{findExtremes}($P$, $n$)${\scriptsize\Comment{{\color{blue} extremos en tiempo $O(n)$}}}
\State Sea $Q[]$ un arreglo de n enteros
\For{\texttt{$j=1$ to $n$}}
	\If{$P[j]$ está fuera del octágono $CP(E)$}
	    \State{\scriptsize\Comment{{\color{blue} encuentra $Q_i$ para $P[j]$ (tiempo $O(1)$)}}}
		\State $i\gets\textsc{findQueue}(P,n,j$)
		\State $Q[j] = i$ 
	\EndIf
\EndFor
\State \textbf{return} $Q$ 
\EndProcedure
\end{algorithmic}
\label{alg_filterGPU}
\end{algorithm}
\end{center}


Como referencia, también se implementó una versión paralela de heaphull en CPU multi-core. Se probaron dos versiones; una con el patrón \textit{reduce} de OpenMP para mínimos y máximos, y otra usando la técnica de \textit{divide and conquer}. Comparaciones de rendimiento mostraron que esta última logra mayor aceleración, de hasta $3x$ para el filtrado, siendo la escogida. 

\section{Evaluación Experimental}
A fin de comparar experimentalmente el desempeño de nuestra solución paralela en GPU, incluimos como baseline las implementaciones de: heaphull tradicional~\cite{ferrada_2019}, CudaChain~\cite{mei_2016} y ConcurrentHull~\cite{masnadi_2020}; este último permite el manejo de una mayor cantidad de puntos y consigue mejores resultados en esas condiciones.
Para las pruebas se utilizó un computador con Procesador Intel\textregistered Core\texttrademark i5-8300H CPU (2.30GHz × 8), 8 GB de memoria RAM, tarjeta gráfica NVIDIA GeForce GTX $1050$ Ti, CUDA Version 11.6 en Ubuntu 20.04 LTS.

\subsection{Tiempos de Ejecución}
Para cada implementación se crean tests puntos de tamaños desde $10^4$ hasta $10^8$. Cada test se ejecuta 100 veces y se calcula el tiempo promedio de ejecución en cada algoritmo.

\begin{figure}[htbp]
\centerline{\includegraphics[width=\columnwidth]{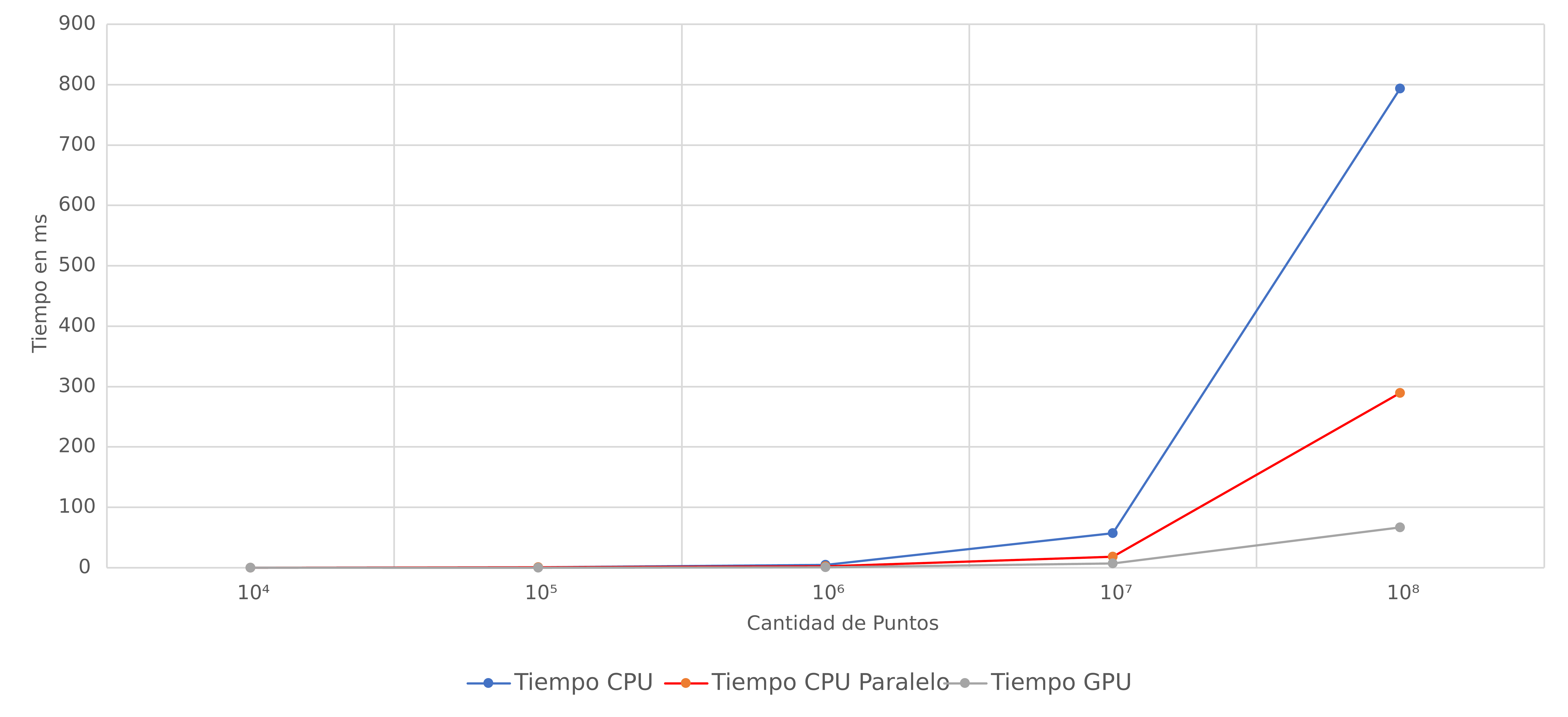}}
\caption{Tiempos promedio de filtrado para las versiones secuencial, paralela en CPU y paralela en GPU.} \label{fig2}
\end{figure}

Los tiempos promedio de búsqueda de puntos extremos para las pruebas experimentales en las versiones de heaphull secuencial, paralelo en CPU y paralelo en GPU se pueden ver en la Tabla~\ref{tabla:tiempos_extremos} y en el gráfico de la Figura~\ref{fig2}. Los resultados de heaphull paralelo CPU solo son una referencia; los gráficos futuros solo usan la versión secuencial (Heaphull CPU).  

\begin{table}[htbp]
\caption{Tiempo promedio de búsqueda de puntos extremos en ms.}
\begin{center}
\begin{tabular}{|c|c|c|c|}
\hline
\textbf{Puntos} & \textbf{Tiempo CPU} & \textbf{Tiempo CPU Paralelo} & \textbf{Tiempo GPU} \\
\hline
$10^4$ & $0.0460$ & $0.1000$ & $0.0433$ \\ \hline
$10^5$ & $0.4600$ & $0.4000$ & $0.0814$ \\ \hline
$10^6$ & $5.0120$ & $2.4000$ & $0.8505$ \\ \hline
$10^7$ & $56.8020$ & $18.0000$ & $6.9785$ \\ \hline
$10^8$ & $793.7420$ & $289.1300$ & $67.0598$ \\ \hline
\end{tabular}
\label{tabla:tiempos_extremos}
\end{center}
\end{table}

\subsubsection{Desempeño caso Promedio (Dist. Normal)}

\begin{figure}[htbp]
\centerline{\includegraphics[width=\columnwidth]{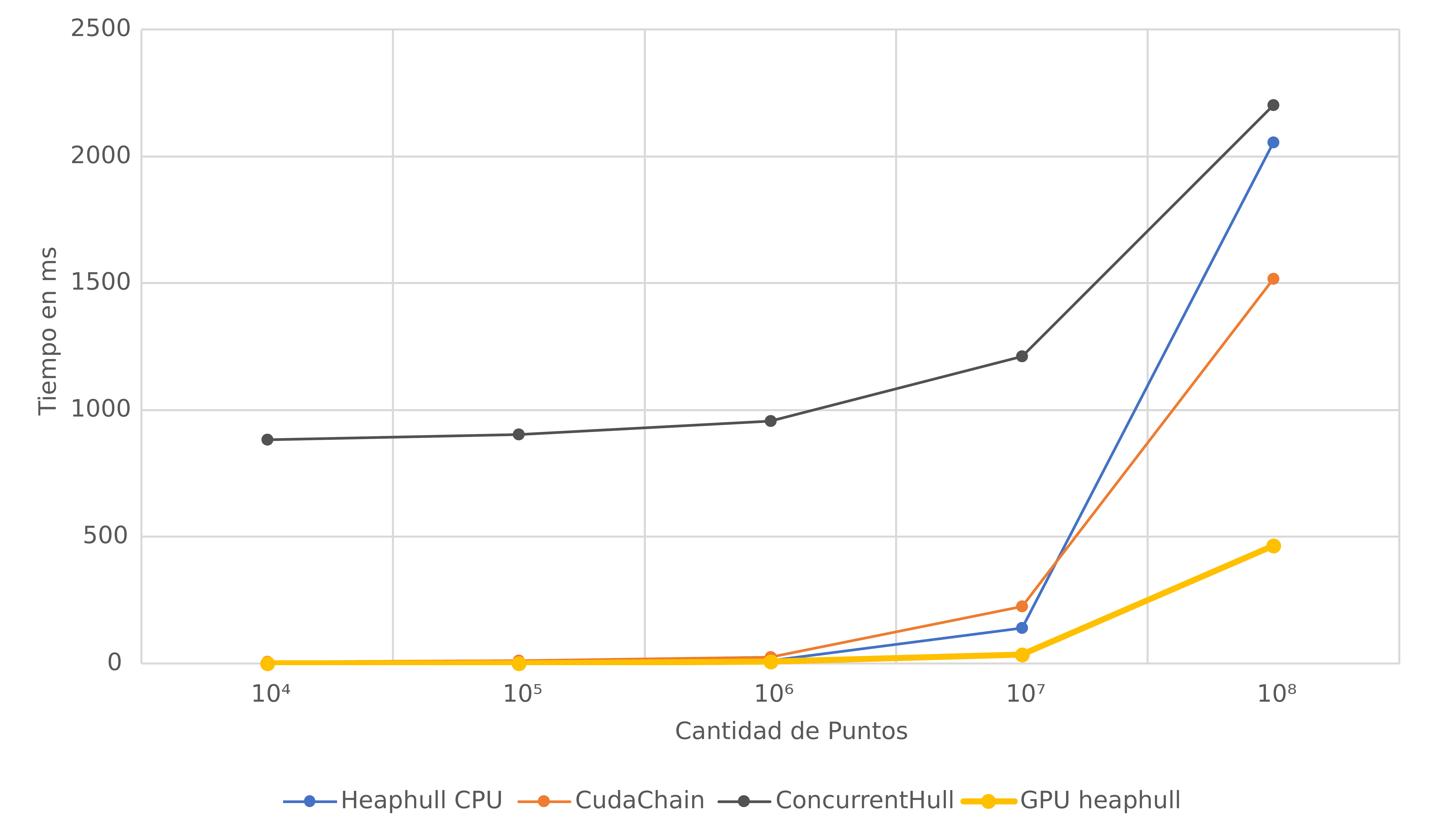}}
\caption{Tiempos para el cómputo del convex hull en el caso promedio} \label{fig3}
\end{figure}

En la Tabla~\ref{tabla:tiempos_ch_gpu} se pueden ver los tiempos de cálculo de la versión secuencial de heaphull, CudaChain, ConcurrentHull y nuestra implementación en GPU, en milisegundos para $10^4$ a $10^8$ puntos. Nuestra solución es la que obtiene mejores tiempos en los casos con $10^5$ puntos o más. En la Figura~\ref{fig3} podemos ver el gráfico de los tiempos en milisegundos para el cálculo de convex hull en las cuatro implementaciones anteriores.

\begin{table}[htbp]
\begin{center}
\caption{Tiempo promedio cómputo del convex hull en el caso promedio.}
\begin{tabular}{|c|c|c|c|c|}
\hline
\textbf{Puntos} & \textbf{Heaphull} & \textbf{CudaChain} & \textbf{ConcurrentHull} & \textbf{GPU HH} \\
\hline
$10^4$ & $0.1201$ & $6.4803$ & $882.0300$ & $0.1575$ \\ \hline
$10^5$ & $1.2014$ & $12.0930$ & $902.0700$ & $0.9940$ \\ \hline
$10^6$ & $12.0824$ & $24.1600$ & $956.3200$ & $7.0058$ \\ \hline
$10^7$ & $139.9730$ & $224.2051$ & $1211.6300$ & $35.1825$ \\ \hline
$10^8$ & $2054.5416$ & $1517.1612$ & $2201.5500$ & $464.8182$ \\ \hline
\end{tabular}
\label{tabla:tiempos_ch_gpu}
\end{center}
\end{table}

\begin{figure}[htbp]
\centerline{\includegraphics[width=\columnwidth]{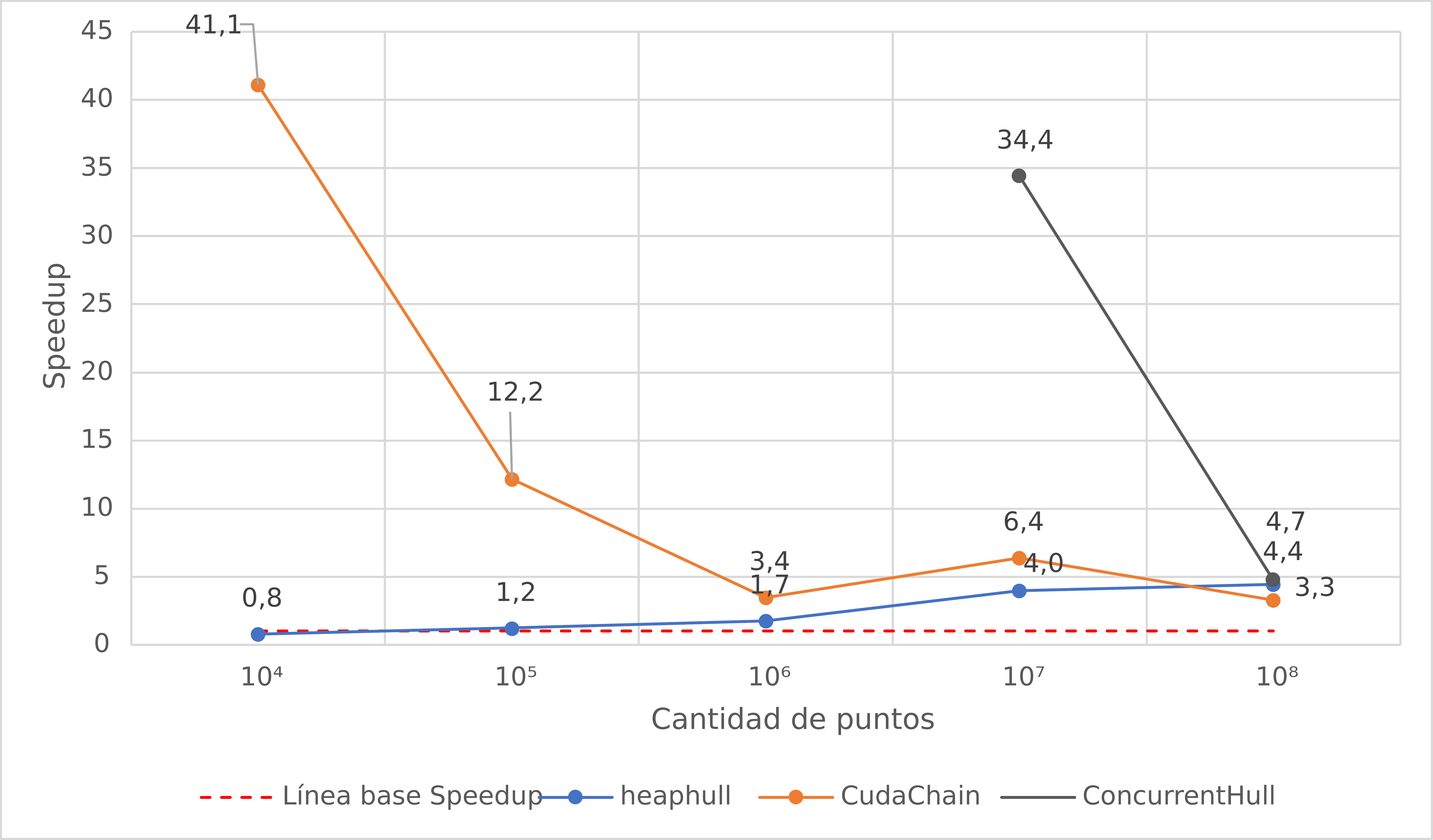}}
\caption{Speedup implementación desarrollada sobre heaphull, CudaChain y ConcurrentHull, caso promedio.} \label{fig5}
\end{figure}

El Speedup obtenido se puede ver en la Tabla~\ref{tabla:speedup} y en la Figura~\ref{fig5}, pudiendo identificarse que  al trabajar con $10^8$ puntos se logra una disminución de tiempo de hasta $4.4\times$ respecto al algoritmo secuencial de heaphull y de $3.2\times$ respecto a CudaChain, el cual ya logra un speedup de hasta $6\times$ comparado con la librería de Quickhull (Qhull en C++). No se considera en el gráfico el speedup frente a ConcurrentHull de menos de $10^7$ puntos para poder mantener una escala apreciable.

\begin{table}[htbp]
\begin{center}
\caption{Speedup solución desarrollada sobre heaphull, CudaChain y ConcurrentHull, caso promedio.}
\begin{tabular}{|c|c|c|c|}
\hline
\textbf{Puntos} & \textbf{Heaphull} & \textbf{CudaChain} & \textbf{ConcurrentHull} \\
\hline
$10^4$ & $0.7625$ & $41.1448$ & $5600,1905$ \\ \hline
$10^5$ & $1.2087$ & $12.1660$ & $907.5151$ \\ \hline
$10^6$ & $1.7246$ & $3.4486$  & $136.5040$ \\ \hline
$10^7$ & $3.9785$ & $3.5303$  & $34.4384$ \\ \hline
$10^8$ & $4.4201$ & $3.2640$  & $4.7364$ \\ \hline
\end{tabular}
\label{tabla:speedup}
\end{center}
\end{table}


En relación al porcentaje de filtrado de puntos, heaphull secuencial y la versión en GPU de este logran porcentajes para puntos provenientes de una distribución normal sobre el 99,99\% en promedio (igual porcentaje para mismos puntos de entrada). En el caso de CudaChain logra un filtrado de 98,86\% en promedio en el mejor caso ($10^8$ puntos de entrada) y en el peor caso para este ($10^4$ puntos de entrada) llega a un 91,6\%, mientras que heaphull en sus dos versiones logra un 99,87\% para esa misma cantidad de puntos.

\subsubsection{Desempeño Peor Caso (Circunferencia)}

\begin{figure}[htbp]
\centerline{\includegraphics[width=\columnwidth]{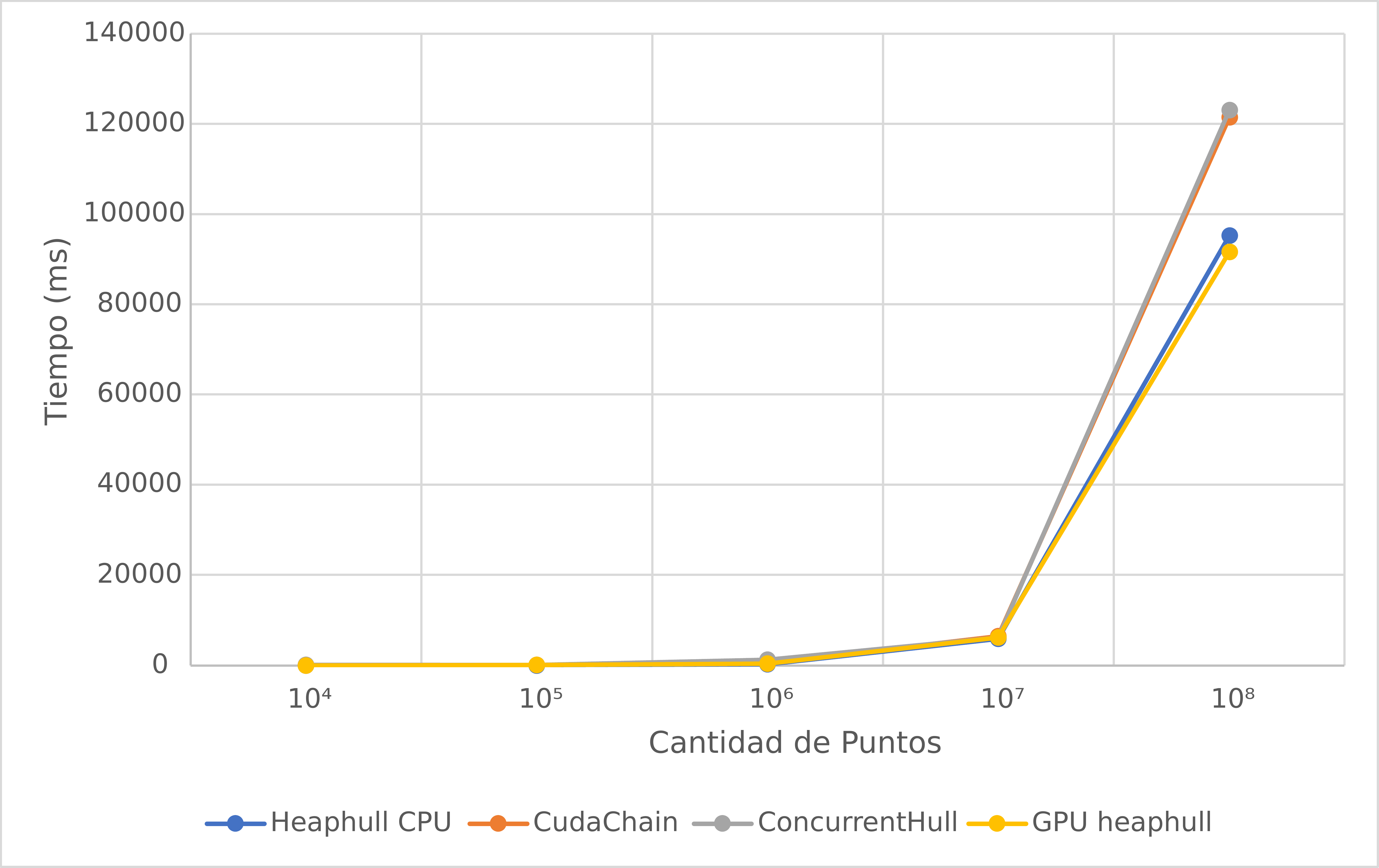}}
\caption{Tiempos de cómputo del convex hull en el peor caso} \label{fig6}
\end{figure}

Para el peor caso (los puntos sobre la circunferencia) con $10^7$ puntos nuestra implementación paralela en GPU demora en encontrar el convex hull $6104.58$ ms, sin embargo con 2\% de distorsión el tiempo disminueye a $4973.3$ ms; el primero representa un speedup de $0.96$ respecto del algoritmo secuencial de heaphull, sin embargo en el segundo caso se obtiene un speedup de $1.41$. Para $10^8$ puntos el speedup es de $1,04$ en el peor caso. En la Tabla~\ref{tabla:tiempos_circunsferencia} se encuentran los tiempos de ejecución en el peor caso. En la figura~\ref{fig6} se puede ver el gráfico con los tiempos de para cada versión.

\begin{table}[htbp]
\begin{center}
\caption{Tiempos para el cómputo del convex hull en el peor caso.}
\begin{tabular}{|c|c|c|c|c|}
\hline
\textbf{Puntos} & \textbf{Heaphull} & \textbf{CudaChain} & \textbf{ConcurrentHull} & \textbf{GPU HH} \\
\hline
$10^4$ & $1.0153$ & $86.7300$ & $90.0223$ & $1.6400$ \\ \hline
$10^5$ & $15.0671$ & $112.5680$ & $96.3250$ & $23.6950$ \\ \hline
$10^6$ & $241.7010$ & $445.2650$  & $1256.3600$ & $427.0750$ \\ \hline
$10^7$ & $5861.8900$ & $6459.5251$  & $6201.5522$ & $6104.5800$ \\ \hline
$10^8$ & $95278.10$ & $121426.00$  & $123045.20$ & $91559.80$ \\ \hline
\end{tabular}
\label{tabla:tiempos_circunsferencia}
\end{center}
\end{table}

Con una distorsión de sólo 2\%, los porcentajes de filtrado para $10^8$ puntos se mantienen en 10.50\% en promedio, con tiempos de ejecución del algoritmo completo similares al caso promedio. La versión codificada del algoritmo tiene un comportamiento con un speedup no relevante cuando se trata de cantidades de puntos inferiores a los $10^7$ puntos. Además, cuando se trata del peor caso no hay una mejora significativa respecto a la versión secuencial.  

En general, los mejores tiempos se logran al trabajar con $10^7$ y $10^8$ puntos, en estos casos el speedup es mayor a 4x.

\section{Discusión y Conclusiones}
Nuestro trabajo presenta un implementación paralela en GPU basada en el algoritmo heaphull que mejora el tiempo de filtrado y el tiempo total de cálculo del convex hull, manteniendo las mismas garantías asintóticas óptimas del algoritmo tradicional para $2D$. El algoritmo se basa en realizar un filtrado con un octágono que permite descartar todos los puntos que se encuentran en su interior y lograr en el caso promedio filtrar el $99,99\%$ de los puntos del conjunto de entrada, quedando una cantidad muy reducida de puntos para el cálculo del convex hull; conservando así la eficiente del filtrado del algoritmo heaphull original.

El mayor impacto en desempeño se logra al trabajar con grandes volúmenes de puntos de entrada (sobre $10^7$), siendo un aporte importante en esta arista para el cálculo del convex hull, lo que es característico en la presente era del Big Data.

En comparación con otras implementaciones en GPU se obtienen mejores tiempos totales, considerando que se está trabajando sobre un algoritmo que ya resulta ser el más rápido secuencial y en $2D$, esto es un importante aporte al desarrollo futuro de implementaciones que puedan requerir el cálculo en tiempo real de grandes volúmenes de puntos.

Dentro de las limitaciones de la solución propuesta se encuentra el uso de memoria de GPU, el cual se ve aumentado por la necesidad de crear arreglos para almacenar los índices que son resultado de las reducciones para poder conservar el arreglo original de puntos de entrada.

Como trabajo futuro tenemos previsto ofrecer un nuevo diseño e implementación del convex hull con algoritmos que sean paralelizables (Jarvis March, Quickhull, Kirkpatrick–Seidel, Chan's algorithm) para comprobar si alguno de ellos, a pesar de aumentar la complejidad del algoritmo, puede lograr mejores tiempos aún al aprovechar el paralelismo de la GPU. Por otra parte, la implementación de la solución para más dimensiones puede ser un importante aporte, por sobre todo en $3D$, pero se debe considerar que el tamaño requerido de memoria crece en potencias según la cantidad de dimensiones, lo cual puede dificultar el cálculo de grandes volúmenes de datos en la GPU.

También en el trabajo futuro está, siguiendo el modelo de heaphull, el uso eficiente de memoria, lo cual podría ser un importante aporte de la mano de grandes volúmenes de datos, logrando data más compacta y facilitando el cálculo de mayores números de datos.

\end{document}